\documentclass[10pt,twocolumn]{IEEEtran}
\usepackage{setspace}

\usepackage{cite}
\usepackage{amsthm}
\usepackage{subcaption}
\usepackage{lmodern} 
\usepackage{amsmath,amssymb, amsfonts}
\usepackage{mathtools}
\usepackage[type1]{libertine}
\usepackage[T1]{fontenc} 
\usepackage[libertine]{newtxmath}
\usepackage{float}
\usepackage{algorithm}
\usepackage{xcolor}

\DeclareSymbolFont{largesymbols}{OMX}{cmex}{m}{n}
\DeclareMathSymbol{\intop}{\mathop}{largesymbols}{"52}
\DeclareMathSymbol{\sumop}{\mathop}{largesymbols}{"50}
\DeclareMathSymbol{\sqrtop}{\mathop}{largesymbols}{"70}
\usepackage{bm}
\allowdisplaybreaks

\usepackage[font=footnotesize]{caption}

\usepackage{algorithmic}
\usepackage{graphicx}
\usepackage{textcomp}
\usepackage[hidelinks, colorlinks=true, linkcolor=blue, urlcolor=blue, citecolor=blue]{hyperref}
\usepackage{cleveref}
\usepackage{commath}
\usepackage{optidef}
\usepackage{orcidlink}
\usepackage{lipsum}
\usepackage{stfloats}
\usepackage{physics}
\usepackage{braket}

\newtheorem{theorem}{Theorem}[section]

\newtheorem{lemma}{Lemma}[section]

\newtheorem{prop}{Proposition}[section]

\begin{document}


\IEEEaftertitletext{\vspace{-1.5\baselineskip}}

\title{\LARGE{QPPG: Quantum-Preconditioned Policy Gradient for Link Adaptation in Rayleigh Fading Channels}}

\author{ Oluwaseyi~Giwa{~\orcidlink{0009-0001-5771-7446},
Muhammad~Ahmed~Mohsin{~\orcidlink{0009-0005-2766-0345}},
Folarin~Jubril~Adesola{~\orcidlink{0009-0007-9934-1081}},
Muhammad~Ali~Jamshed{~\orcidlink{0000-0002-2141-9025}}
}

\thanks{Oluwaseyi Giwa is with the African Institute for Mathematical Sciences, South Africa (e-mail: \{oluwaseyi\}@aims.ac.za).\newline
Muhammad Ahmed Mohsin is with Stanford University, Stanford, California, 94305, United States (e-mail: \{muahmed\}@stanford.edu).\newline
Folarin Adesola is affiliated with the Department of Electrical Engineering at Olabisi Onabanjo University, Nigeria (e-mail:\{jubrilfolarin8\}@gmail.com). \newline
M. A. Jamshed is with the College of Science and Engineering, University of Glasgow, UK (e-mail: \{muhammadali.jamshed\}@glasgow.ac.uk).}
\thanks{codebase: \url{https://github.com/OluwaseyiWater/qppg-refactor}}
}

\makeatletter
\patchcmd{\@maketitle}
{\addvspace{0.5\baselineskip}\egroup}
{\addvspace{0\baselineskip}\egroup}
{}
{}
\makeatother

\maketitle
\begin{abstract}
Reliable link adaptation is critical for efficient wireless communications in dynamic fading environments. However, reinforcement learning (RL) solutions often suffer from unstable convergence due to poorly conditioned policy gradients, hindering their practical application. We propose the quantum-preconditioned policy gradient (QPPG) algorithm, which leverages Fisher-information-based preconditioning to stabilise and accelerate policy updates. Evaluations in Rayleigh fading scenarios show that QPPG achieves faster convergence, a \(28.6\%\) increase in average throughput, and a \(43.8\%\) decrease in average transmit power compared to classical methods. This work introduces quantum-geometric conditioning to link adaptation, marking a significant advance in developing robust, quantum-inspired reinforcement learning for future 6G networks, thereby enhancing communication reliability and energy efficiency.
\end{abstract}

\begin{IEEEkeywords}
Link adaptation, quantum-preconditioned policy gradient (QPPG), 6G networks, fading channels.
    \end{IEEEkeywords}

\section{Introduction}\label{sec:introduction}
\IEEEPARstart{W}{ireless} communication over fading channels remains one of the fundamental challenges in modern networks. In particular, Rayleigh fading channels, which model rich-scattering non-line-of-sight environments, cause rapid and unpredictable fluctuations in signal strength that can significantly degrade throughput and reliability. To mitigate these effects, link adaptation techniques such as adaptive modulation and coding (AMC) and power control have been extensively studied as key enablers of efficient spectrum use~\cite{goldsmith2005, gross2006}. Early works on link adaptation for Rayleigh fading channels demonstrated how explicit channel estimation and threshold-based switching could improve throughput and maintain robustness under fading conditions~\cite{taek2001, ismail2004, tran2012, bae2013}.

Despite their success, these classical approaches rely on accurate channel estimation, fixed rules, and often compromise between average throughput and outage probability in a suboptimal manner~\cite{ismail2004, tran2012, bae2013}. Furthermore, as networks evolve toward 6G with denser topologies and stringent reliability demands, such schemes struggle to scale or adapt to system-level complexities~\cite{walid2020, mohsin2025}. Recent works have explored deep reinforcement learning (DRL) and meta reinforcement learning (RL) for link adaptation and resource allocation, showing promising adaptability but still facing high sample complexity and training instability~\cite{sungho2010, hierdrl2025, giwa2025, meta2025}.

In this letter, we propose quantum-preconditioned policy gradient (QPPG), a natural actor-critic method for link adaptation in Rayleigh fading channels. QPPG integrates RL with quantum-inspired preconditioning, using Fisher information geometry to stabilise policy updates. By combining natural gradient principles with quantum Fisher preconditioning, QPPG efficiently navigates the non-convex optimisation landscape of policy learning while retaining scalability to continuous action spaces such as transmit power control.

Our key contributions are that we model link adaptation as a partially observable Markov decision process (POMDP) with latent fading states, noisy pilot-based observations, and joint modulation and coding schemes (MCS) and power control actions. We then design the QPPG framework, integrating Fisher-preconditioned policy updates with a critic baseline for stable training in fading environments. Furthermore, we provide theoretical insights into convergence and complexity, demonstrating how quantum preconditioning enhances gradient conditioning. Using five representative network scenarios, we benchmark QPPG against classical natural policy gradient (NPG)~\cite{amari1998, kakade2001} and quantum actor-critic (QAC).
Results averaged over ten seeds and several network scenarios demonstrate a \(28.6\%\) increase in throughput and \(43.8\%\) decrease in transmit power for QPPG over QAC and NPG.


\section{Problem Formulation}\label{sec:problem-formulation}
\subsection{Assumptions}\label{sec:assumptions}
Throughout this work, we adopt several assumptions to simplify the formulation and isolate the impact of link adaptation. We consider an \textbf{i.i.d. block fading} model where the multiple-input multiple-output (MIMO) channel vector \(h_{t}\in\mathbb{C}^{N}\) is regenerated independently at the beginning of each coherence block, with distribution \(h_{t}\sim\mathcal{CN}(0,I_{N})\), implying no temporal correlation. The receiver noise variance is assumed to be \textbf{constant} \(\sigma^{2}\) across all blocks. At the start of each block, the transmitter performs \textbf{pilot-based estimation} to obtain a channel noisy estimate \(\hat{h}_{t}\) with an effective signal-to-noise ratio (SNR) \(\mathrm{SNR}_{\text{pilot}}\). The corresponding estimation noise follows \(\varepsilon_{t}\sim\mathcal{CN}(0,\frac{\sigma^{2}}{\mathrm{SNR}_{\text{pilot}}} I_{N})\). To account for imperfect receiver calibration, we introduce a \textbf{noise-uncertainty model} where the estimated noise variance is perturbed as \(\hat{\sigma}_{t}^{2}=\sigma^{2}\cdot 10^{U_{t}/10}\), with the uncertainty term \(U_{t}\sim\mathrm{Unif}[-\Delta_{\mathrm{dB}},\Delta_{\mathrm{dB}}]\). The available \textbf{action space} comprises adaptive modulation choices from the set \(m_{t} \in \{4,16,64\}\)-QAM and a continuous transmit power \(p_{t}\in[0,P_{\max}]\). As coding rates are not explicitly modelled, the throughput in bits per symbol is given by \(r_{\text{th}}(m)=\log_{2}M(m)\).

Under these assumptions, the problem reduces to a POMDP with fresh state realisation each block.

\subsection{Partially Observable Markov Decision Process}\label{sec:pomdp}
Given the assumptions in Section~\ref{sec:assumptions}, we formalise the link adaptation problem as a POMDP.
\begin{equation}\label{eq:pomdp}
    \mathcal{P} = \left(\mathcal{S, \; A, \; O,} \; T, \; \Omega, \; R, \; \gamma\right),
\end{equation}
with latent channel state, partial observations via pilots, and control over modulation and power.\\
\textbf{Latent state \(\mathcal{S}\):} At discrete time \(t\) (one block), the hidden state is:
\begin{equation}\label{eq:hidden-state}
    s_{t} \triangleq \left(h_{t}, \; \sigma^{2}\right), \quad \sigma^{2} > 0
\end{equation}

The transition kernel is defined by the i.i.d. block fading law, 
    \begin{equation}\label{eq:transition-kernel}
        T \left(s_{t + 1} | s_{t}\right) = f_{h}(h_{t+1}) \cdot \delta(\sigma_{t+1}^{2} -\sigma^{2})
    \end{equation}
where \(f_{h}\) is the circularly symmetric complex Gaussian Rayleigh probability density function (pdf) and \(\delta\) enforces constant noise power.\\
\textbf{Actions \(\mathcal{A}\):} Each action selects a modulation/coding mode and a continuous transmit power (Section~\ref{sec:assumptions}):
\begin{equation}\label{eq:action-space}
    a_{t} \triangleq \left(m_{t}, \; p_{t}\right),\\
\end{equation}

\textbf{Observations \(\mathcal{O}\):} After pilots, the agent receives a noisy channel estimate
\begin{equation}\label{eq:observation}
    o_{t} \triangleq \left(\Re \{\hat{h}_{t}\},\; \Im\{\hat{h}_{t}\}, \; \hat{\sigma}_{t}^{2}\right) \in \mathbb{R}^{2N + 1}
\end{equation}
\textbf{Observation model:} With pilot SNR, \(\text{SNR}_{\text{pilot}}\) and true noise \(\sigma^{2}\),
\begin{align}
    \hat{h}_{t} = h_{t} + \varepsilon_{t},\\
    \varepsilon_{t} \sim \mathcal{CN}\left(0, \frac{\sigma^{2}}{\text{SNR}_{\text{pilot}}} I_{N}\right)
    \end{align}
and an estimated noise variance (Section~\ref{sec:assumptions}). This defines \(\Omega (o_{t}|s_{t})\)\\
\textbf{Reward \(R\):} Let the instantaneous post-combining SNR for action \(a_{t} = (m_{t}, \; p_{t})\)
\begin{equation}\label{eq:instantaneous-post}
    \gamma_{t} = \frac{p_{t}\left|\left|h_{t}\right|\right|_{2}^{2}}{\sigma^{2}}
\end{equation}
A simple success/threshold model uses modulation-dependent thresholds \(\gamma_{\text{thr}}(m)\):
\begin{equation}\label{eq:thereshold}
    \text{succ}_{t} = 1 \left\{\gamma_{t} \geq \gamma_{\text{thr}}(m_{t})\right\}
\end{equation}
With linear power penalty \(\lambda \in [0, 1)\), the reward becomes:
\begin{equation}\label{eq:reward}
    R(s_{t}, a_{t}) = \text{succ}_{t} \; \cdot \; r_{\text{th}}(m_t) \; \cdot \; \left(1 - \lambda\frac{p_t}{\text{P}_{\text{max}}}\right) - \left(1 - \text{succ}_{t}\right)\text{c}_{\text{fail}},
\end{equation}
where \(\text{c}_{\text{fail}} > 0\) is a small penalty for failed transmissions. The observed one-step reward is \(r_{t} = R(s_{t}, a_{t})\).

\section{Quantum-Preconditioned Policy Gradient }\label{sec:qppg}
\begin{figure}[t]
    \centering
    \includegraphics[width=\linewidth]{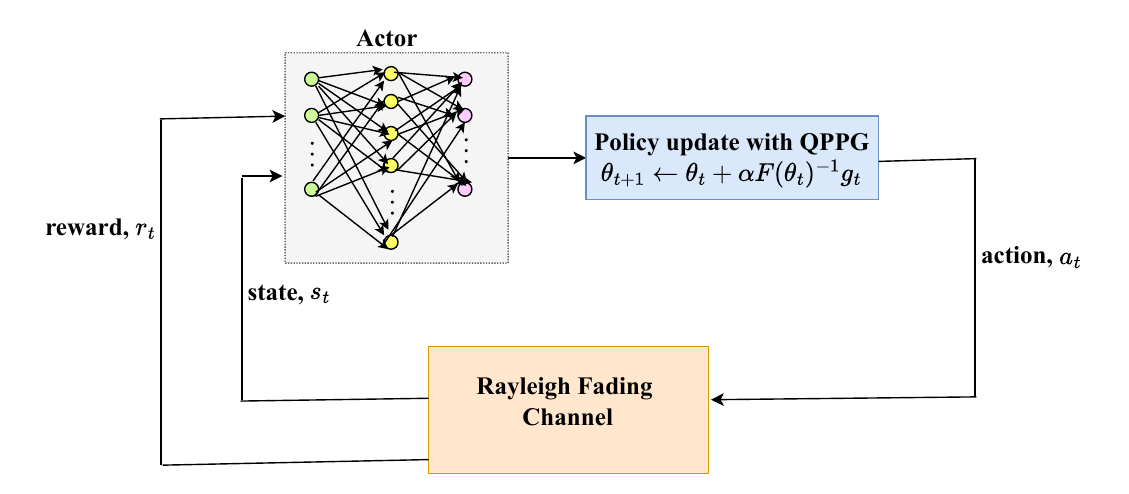}
    \caption{QPPG framework for link adaptation over Rayleigh fading channels.}
    \label{fig:qppg-framework}
\end{figure}

The central challenge in policy gradient RL for link adaptation is the slow and unstable convergence of vanilla gradient updates, particularly in high-dimensional parameter spaces. QPPG method addresses this by introducing a Fisher-information-based preconditioning step, approximating the \emph{natural gradient}. In classical NPG, the update direction requires solving a linear system involving the Fisher information matrix (FIM), which is computationally prohibitive for large networks. QPPG introduces a scalable approximation motivated by quantum linear system solvers, which can be implemented entirely in classical simulation. It combines:
\begin{itemize}
    \item an \emph{actor} that outputs modulation and power distributions,
    \item a \emph{critic} that estimates state values for variance reduction,
    \item a conjugate gradient solver for approximating \(F^{-1}g\), where \(g\) is the vanilla policy gradient and \(F\) the Fisher matrix.
\end{itemize}
Fig.~\ref{fig:qppg-framework} shows a simplified scenario of the QPPG algorithm acting in the Rayleigh channel.

\subsection{Mathematical Formulation}\label{sec:mathematical-formulation}
Let \(\pi_{\theta}(a|s)\) denote the stochastic policy parameterised by \(\theta\), where \(a=(m,p)\) encodes modulation order and power, and \(s\) is the observation vector defined in Section~\ref{sec:problem-formulation}.  
The policy gradient is:
\begin{equation}\label{eq:policy-gradient}
    g \coloneqq \nabla_{\theta} J(\theta) = \mathbb{E}_{\pi_{\theta}}\left[ A(s,a) \nabla_{\theta}\log \pi_{\theta}(a|s)\right],
\end{equation}
where \(A(s, a)\) is the advantage function, estimated via Generalised Advantage Estimation (GAE).

The natural gradient update requires:
\begin{equation}\label{eq:natural-gradient}
    \theta_{t+1} = \theta_{t} + \alpha F(\theta_{t})^{-1} g_{t},
\end{equation}
with FIM
\begin{equation}\label{eq:fisher-matrix}
    F(\theta) = \mathbb{E}_{\pi_{\theta}}\left[ \nabla_{\theta} \log \pi_{\theta}(a|s) \nabla_{\theta} \log \pi_{\theta}(a|s)^{\top}\right].
\end{equation}

QPPG avoids explicit inversion of \(F\) by solving
\begin{equation}
    F x = g
\end{equation}
via conjugate gradient, with Fisher-vector products (FVP) approximated on sampled trajectories:
\begin{equation}
    \text{FVP}(v) = \frac{1}{M}\sum_{i=1}^{M} g_{i}\left(g_{i}^{\top}v\right) + \xi v,
\end{equation}
where \(g_{i}\) is the per-sample score function gradient, \(\xi>0\) is a damping factor, and \(M\) is the subsample size. Algorithm~\ref{alg:qppg} summarises the training loop.

\begin{algorithm}[t]
\caption{QPPG for Link Adaptation in Rayleigh Fading Channel}
\label{alg:qppg}
\begin{algorithmic}[1]
\STATE \textbf{Initialise} policy parameters \(\theta\), critic parameters \(\phi\).
\FOR{episodes \(=\;1 \; \text{to} \; E\)}
    \STATE Collect trajectory \(\tau = \{(s_{t},a_{t},r_{t})\}\).
    \STATE Estimate advantages $A_{t}$ via GAE with \(V_{\phi}(s)\).
    \STATE \(g = \nabla_{\theta} \sum_{t}\log\pi_{\theta}(a_{t}|s_{t}) A_{t}\).
    \STATE Approximate \(x = F^{-1} g\) using conjugate gradient and FVP.
    \STATE \(\theta_{t} \leftarrow \theta_{t} + \alpha F(\theta_{t})^{-1}g_{t}\).
    \STATE Update critic by minimising \((V_{\phi}(s_{t}) - R_{t})^{2}\).
\ENDFOR
\end{algorithmic}
\end{algorithm}

\subsection{Theoretical Analysis}\label{sec:theoretical-analysis}
We provide a sketch of the convergence properties.

\begin{lemma}[Fisher Positivity]
The FIM, \(F(\theta)\), is symmetric positive definite (SPD), ensuring well-posedness of the linear system \(F x = g\) with damping \(\xi>0\).
\end{lemma}

\begin{prop}[Convergence of Conjugate Gradient]
Given SPD \(F\), conjugate gradient converges to the exact solution in at most \(d\) steps for \(d\) parameters. With damping \(\xi\), the approximate solution \(x\) satisfies
\begin{equation}
    \|x - F^{-1}g\|_{F} \leq \epsilon,
\end{equation}
for tolerance \(\epsilon\) determined by the conjugate gradient iterations.
\end{prop}

\begin{theorem}[Policy Improvement under Natural Gradient]
Under standard assumptions of bounded rewards and sufficient exploration, the QPPG update guarantees monotonic improvement in expected return up to approximation error in the conjugate gradient step.
\end{theorem}

\textbf{Complexity:} Each episode involves (i) actor and critic forward/backward passes \(O(d)\), (ii) FVP evaluations \(O(Md)\), and (iii) conjugate gradient iterations \(O(\text{conjugate gradient}\times Md)\). In practice, this is more expensive than vanilla policy gradient but converges in fewer episodes, reducing sample complexity. To ground this theoretical discussion, we perform our experiments on an NVIDIA Tesla P100 GPU. On average, a single policy update for NPG required \(\approx 35 \;ms\), while QPPG took \(\approx 65\; ms\). This confirms the expected computational hierarchy and highlights the core trade-off: a higher per-step cost for improved sample efficiency.

\section{Experimental Setup}\label{sec:experiment-setup}
To evaluate the proposed QPPG against NPG and QAC, we design a set of experiments in a simulated Rayleigh block-fading channel environment. Each experiment corresponds to a distinct network scenario capturing a key challenge in practical link adaptation. 

\subsection{Network Scenarios}\label{sec:network-scenarios}
We consider five representative scenarios, denoted as \(s1 - s5\), each parameterised by the number of antennas (\(N\)), pilot SNR, and the degree of channel uncertainty (\(\Delta_{dB}\)). Scenario \(s1\) (Baseline) is a standard setting with \(N = 4\), \(\text{SNR}_{\text{pilot}} = 10 \; dB\), and \(\Delta_{dB} = 0\). This scenario provides a controlled environment to benchmark learning stability and steady-state performance. Scenario \(s2\) (High-Dimensional Channels) has \(N=8\), \(\text{SNR}_{\text{pilot}} = 10\;dB\), and \(\Delta_{dB} = 0\). This setting stresses the scalability of the learning algorithms with larger state dimensions. Scenario \(s3\) (Low-Quality CSI) consists of \(N=4\) and \(\text{SNR}_{\text{pilot}} = 10 \; dB\). This scenario simulates degraded channel estimates due to poor training signal quality. It emphasises the robustness of the agent to inaccurate channel state information (CSI). Scenario \(s4\) (Noise Uncertainty) has \(N=4\) and \(\text{SNR}_{\text{pilot}} = 10 \; dB\), but the noise variance estimate is perturbed by an uncertainty window of \(\Delta_{dB}=5\). This represents practical scenarios where the noise floor is not precisely known at the transmitter. Scenario \(s5\) (Combined Challenge) is a more challenging setting with \(N=8\), \(\text{SNR}_{\text{pilot}} = 10 \; dB\), and \(\Delta_{dB}=5\). This combines both s2 and s4, serving as the most demanding benchmark.

\subsection{Performance Metrics}\label{sec:performance-metrics}
To assess learning performance and communication efficiency, we track the physical-layer metrics.\newline
\textbf{Average Throughput:} The mean number of successfully transmitted bits per channel use, reflecting system-level data rate.\newline
\textbf{Power Efficiency:} Average transmit power relative to \(P_{\text{max}}\), characterising energy efficiency.\newline
\textbf{Average Packet Error Rate (PER):} The fraction of failed transmissions, indicating reliability of the learned link adaptation strategy.\newline
This metric evaluation ensures that QPPG is assessed not on abstract learning curves, but on communication-theoretic metrics that determine its practical viability in wireless systems.

\section{Results}\label{sec:results}
\subsection{Physical-Layer Metrics}\label{sec:physical-layer-metrics}
\begin{figure*}[t]
    \centering
    \begin{subfigure}{0.32\textwidth}
        \includegraphics[width=\textwidth]{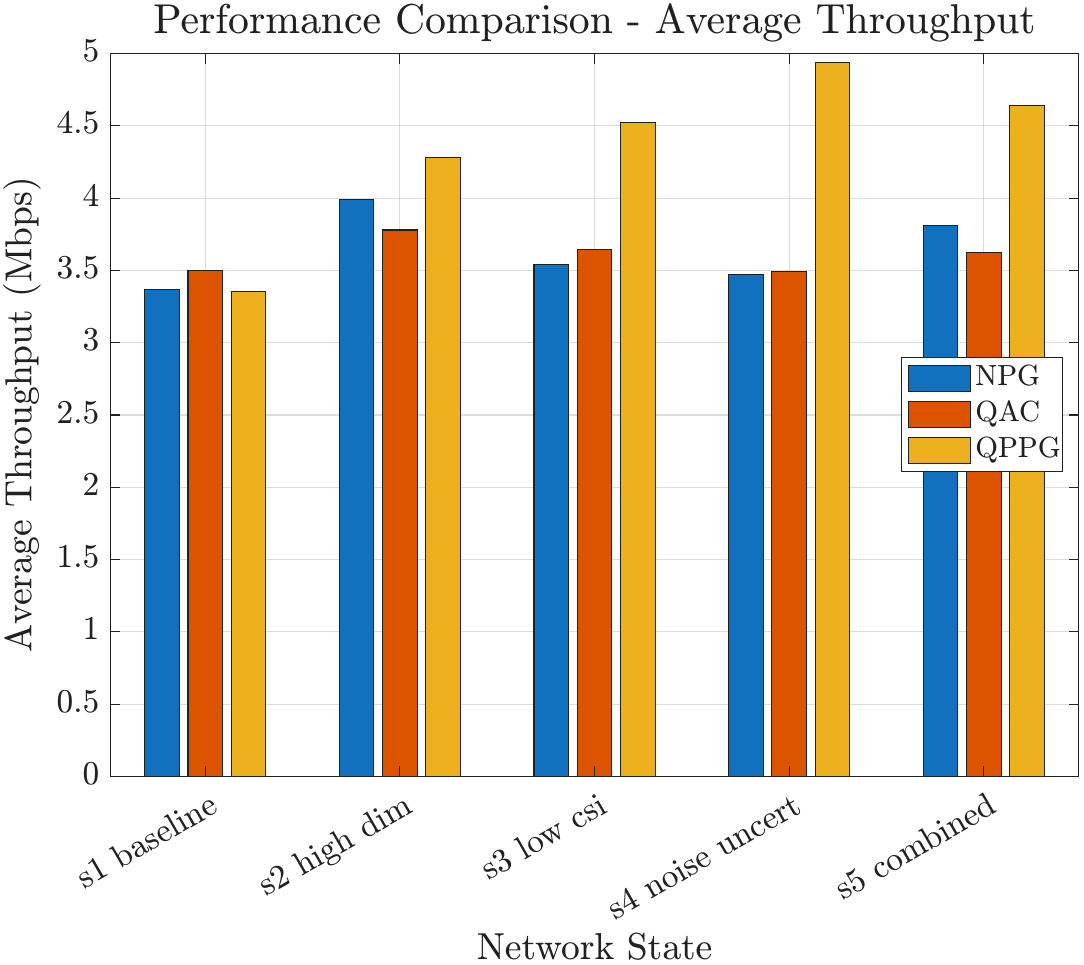}
        \caption{}
        \label{fig:throughput}
    \end{subfigure}
    \begin{subfigure}{0.32\textwidth}
        \centering
        \includegraphics[width=\textwidth]{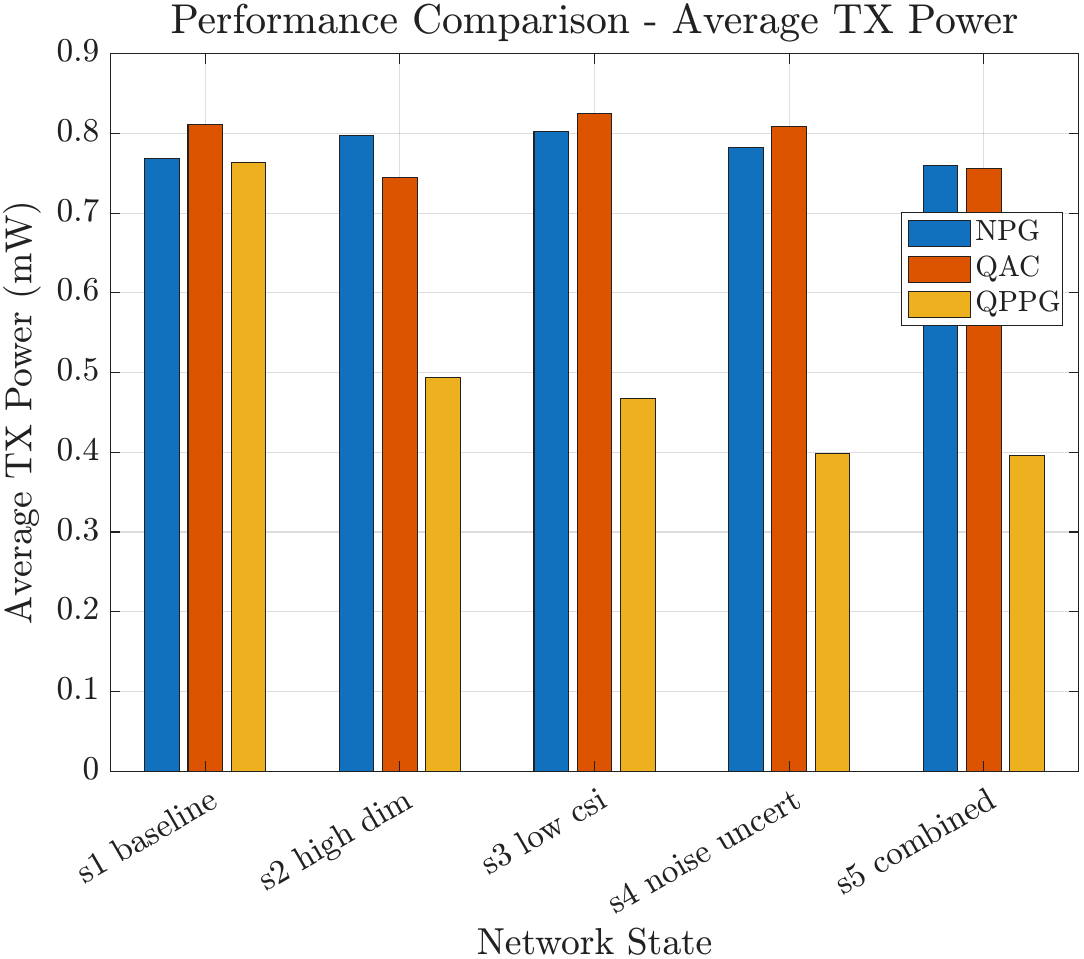}
        \caption{}
        \label{fig:power}
    \end{subfigure}
    \begin{subfigure}{0.32\textwidth}
        \centering
        \includegraphics[width=\textwidth]{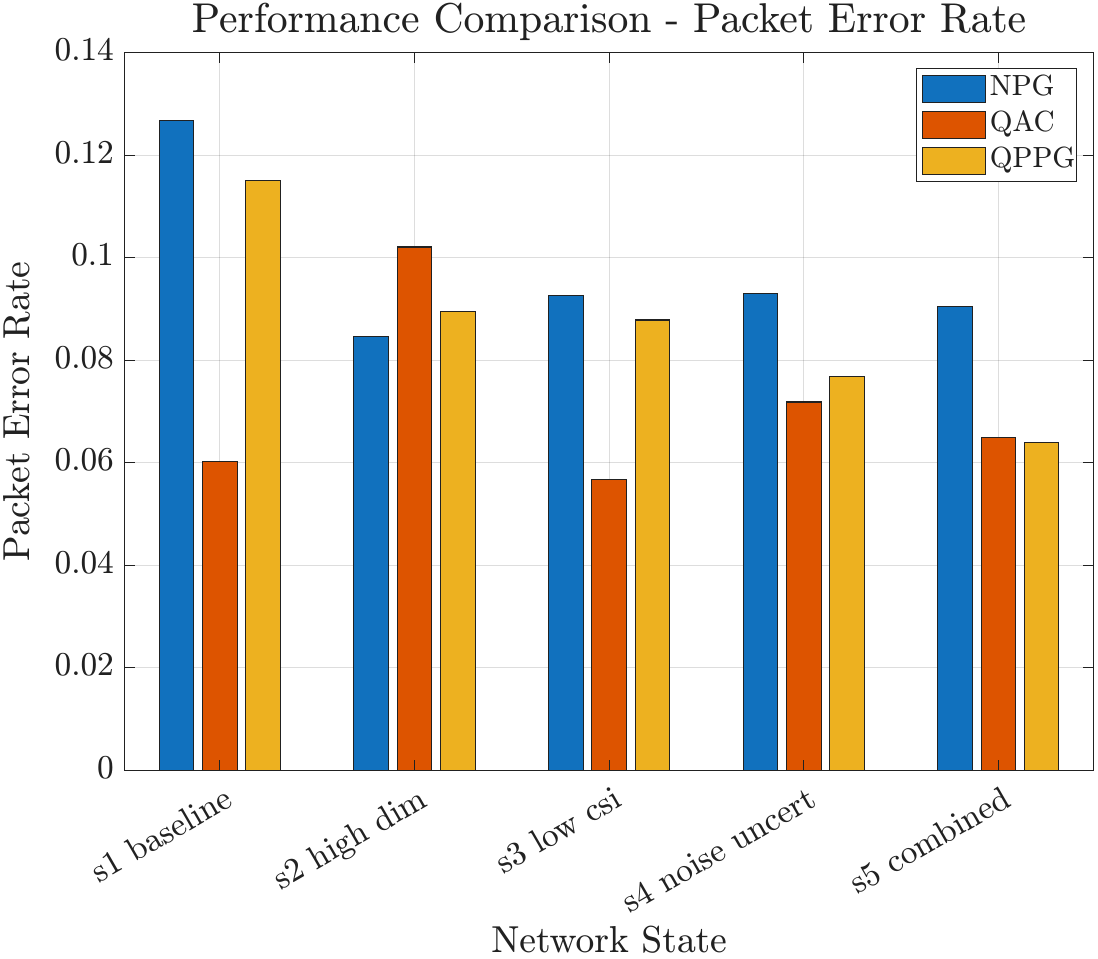}
        \caption{}
        \label{fig:per}
    \end{subfigure}
    \caption{Physical-layer performance metrics for NPG, QAC, and QPPG. \textbf{(a)} Average throughput (Mbps) across network scenarios (higher value is better). \textbf{(b)} Average transmit power across network scenarios (lower value is better). \textbf{(c)} Packet error rate across network scenarios (lower value is better).}
    \label{fig:metrics_plots}
\end{figure*}

Fig.~\ref{fig:metrics_plots} shows the physical-layer metrics results for NPG, QAC, and QPPG algorithms. QPPG achieves higher throughput and lower transmit power than NPG and QAC. This indicates that the quantum-preconditioned policy efficiently utilises available power resources, a key requirement for energy-constrained systems. The PER plot shows that QPPG maintains a moderate error rate, lower than NPG but higher than QAC in some network scenarios. This trade-off highlights that while QPPG improves spectral efficiency through its quantum-inspired preconditioning, it may occasionally select aggressive MCS-power pairs in marginal SNR regions.

Overall, QPPG balances spectral efficiency and power economy better than NPG, demonstrating its potential for sustainable link adaptation.

\subsection{Ablation Study}\label{sec:ablation-study}
We assess the sensitivity of QPPG to various damping factors (\(\xi\))\textemdash which regulates the conditioning strength in the quantum Fisher-based preconditioner\textemdash and network conditions. Table~\ref{tab:xi_ablation} shows the quantitative result for the damping factor. A clear trend emerges: performance improves with increasing \(\xi\), stabilising around \(\xi = 0.5 - 1.0\). For minimal damping (\(\xi < 0.1\)), instability arises due to near-singular Fisher estimates, resulting in noisy policy gradients and lower rewards.
Thus, \(\xi = 0.5\) provides a favourable trade-off between convergence speed and robustness, consistent with theoretical expectations from preconditioned optimisation methods.

\begin{table}[t]
\centering
\caption{Ablation Study on Damping Factor (\(\xi\)) over 10 seeds for 1000 episodes}
\label{tab:xi_ablation}
\begin{tabular}{|c|c|}
\hline
\textbf{Damping Factor (\(\xi\))} & \textbf{Final Avg. Reward} \\
\hline
0.001 & \(47.25 \pm 9.77 \) \\
0.010 & \(42.67 \pm 12.64\) \\
0.100 & \(62.19 \pm 9.71\) \\
0.500 & \(64.99 \pm 8.54\) \\
1.000 & \(65.78 \pm 6.81\) \\
\hline
\end{tabular}
\end{table}


\section{Conclusion}\label{sec:conclusion}
We introduced QPPG, a quantum-preconditioned RL framework for link adaptation in Rayleigh fading channels. By modelling link adaptation as a POMDP and using Fisher-based quantum preconditioning, QPPG stabilises learning and improves convergence without increasing model complexity. Across diverse channel conditions, QPPG achieved a higher average throughput and lower transmit power than NPG and QAC baselines. These results highlight the potential of quantum-geometric optimisation for data-efficient reinforcement learning in wireless systems. Future work will extend QPPG to multi-user settings and explore hybrid quantum-classical implementations for real-time link adaptation in 6G networks.

\bibliographystyle{IEEEtran}
\bibliography{references/references}

\end{document}